\documentclass[aps,prd,preprint,superscriptaddress,showpacs,nofootinbib]{revtex4}
\usepackage{epsf,epsfig,graphics,graphicx}
\usepackage{verbatim,color,ulem}
\newcommand{\be}{\begin{equation}}
\newcommand{\ee}{\end{equation}}
\newcommand{\bea}{\begin{eqnarray}}
\newcommand{\eea}{\end{eqnarray}}
\newcommand{\ba}{\begin{array}}
\newcommand{\ea}{\end{array}}
\usepackage{amssymb,amsmath,amsthm,graphicx}
\usepackage[mathscr]{eucal}
\usepackage{enumerate,color,verbatim,multirow,comment}

\begin{document}

\title{Time dependent field correlators from holographic EPR pairs}
\author{Shoichi Kawamoto}
\email{kawamoto_s@obirin.ac.jp}
\affiliation{Department of Physics,
National Tsing Hua University, Hsinchu, Taiwan, R.O.C.}
\affiliation{Center for High Energy Physics,
Chung Yuan Christian University, Taoyuan, Taiwan, R.O.C.}
\affiliation{College of Arts and Sciences,
J.~F.~Oberlin University, Tokyo, Japan}
\author{Da-Shin Lee}
\email{dslee@gms.ndhu.edu.tw} \affiliation{Department of Physics,
National Dong Hwa University, Hualien, Taiwan, R.O.C.}
\author{Chen-Pin Yeh}
\email{chenpinyeh@gms.ndhu.edu.tw} \affiliation{Department of
Physics, National Dong Hwa University, Hualien, Taiwan, R.O.C.}

\begin{abstract}
 We study the correlators of the fields that couple to the quark and anti-quark EPR pair in the super Yang-Mills theory using the holographic description, which is a string in AdS space with its two ends anchoring on the boundaries. We consider the cases that the endpoints of the string are static and that the endpoints are uniformly accelerated in opposite directions where the exact solutions for the string's profiles are available. In both cases, the two-point correlators of the boundary field, described by the linearized perturbations in the worldsheet, can also be derived exactly where we obtain the all-time evolution of the correlators. In the case of the accelerating string, the induced geometry on the string worldsheet has the causal structure of a two-sided AdS black hole with a wormhole connecting two causally disconnected boundaries, which can be a realization of the ER=EPR conjecture. We find that causality plays a crucial role in determining the nature of the dispersion relation of the particle and the feature of the induced mutual interaction between two particles from the field. In the case that two boundaries of the worldsheet are causally disconnected, the induced effect from the field gives the dissipative dynamics of each particle with no dependence on the distance between two particles, and the induced mutual coupling between them vanishes in the late times, following a power law. When two ends are causally connected, the induced dispersion relation becomes non-dissipative in the late times. Here, we will also comment on the implications of our findings to the entangled particle dynamics and the ER=EPR conjecture.

\end{abstract}

\maketitle

\section{introduction}
Time-dependent field correlators provide a qualitative description of how fast the information spreads out in a quantum theory. This is also related to how fast two entangled subsystems get disentangled. In the content of the AdS/CFT correspondence, ER = EPR is a conjecture stating that two entangled particles, the so-called Einstein-Podolsky-Rosen or EPR pair, can be associated in a gravity theory by a wormhole background called Einstein-Rosen or ER bridge \cite{ER=EPR}. One realization is the eternal AdS black hole, which is dual to the thermofield double state \cite{Maldacena_01}. In this case, the correlators for the operators separately inserted at two causally disconnected boundaries decay exponentially in the boundary time, which is in tension with the information conservation for a quantum black hole \cite{Maldacena_01}. Another realization of the ER=EPR conjecture is a model consisting of a string in AdS space with its two ends located at the AdS boundary, which is dual to the entangled pair of the quark and anti-quark in the super Yang-Mills theory. As the two ends are uniformly accelerating in the opposite directions, the exact worldsheet profile was found in \cite{Xiao_08} and had the same causal structure as the eternal AdS black hole. From this exact solution, it was proposed in \cite{Jensen_13} and \cite{Sonner_13} that the EPR pair of the quark and anti-quark is holographically encoded in the induced wormhole geometry on the string worldsheet. For example, the entanglement entropy of the EPR pair is dual to the entropy in the two-sided black hole of the worldsheet, and the gravity action is dual to the EPR pair production rate.

How fast the information spreads out is also crucial in quantum information processing. The viability of quantum computation depends on how long we can maintain the system in coherent or entangled states without losing information to environments. Thus, it is interesting to study how quantum fields influence the entanglement of the system. We may start with the initial density matrix of the system and environment. The effective theory for the system is described by the reduced density matrix, which is obtained by tracing out environmental variables in the full density matrix. It is an essential quantity for studying the dynamics of entanglement \cite{WC}. The references \cite{SY1,SY2} study the entanglement of two objects interacting with the common free quantum field when two objects either undergo uniform acceleration or are at rest. They found that the entanglement between two objects can be created through the coupling with the quantum field and then decays to vanishing  in some finite duration (called the ``sudden death'' of quantum entanglement \cite{YU}).

The idea of the holographic duality has also been applied to study strong-coupling problems in condensed matter systems and the hydrodynamics of the quark-gluon plasma (see~\cite{Hartnoll_09,Rangmanai_09} for reviews). There are considerable efforts to employ the holographic duality to explore the dissipation behavior of a particle moving in a strongly coupled environment. In these studies, the string's endpoint on the boundary of the AdS black hole serves as a probe particle. The reference \cite{Holographic QBM} reviews the application to non-equilibrium Brownian motion. We have employed this approach to study various behaviors of Brownian particles in the strongly coupled fields, which  potentially can be verified experimentally \cite{Lee_13,Lee_15, Lee_16}.

In this work, we would like to adopt the holographic approach to obtain the time-dependent correlators of the strongly coupled fields that couple to two entangled particles of either undergoing uniform acceleration or being static. The dual gravity description is a probe string in AdS space with its two ends anchoring at the boundary. In the case of the uniform accelerating pair, the setting is the same as in \cite{Xiao_08}. In both cases, we obtain the exact time-dependent field correlators by solving the linearized equation of motion for the perturbations of the string Nambu-Goto actions. Compared to the one-end string, the solutions here are fixed by the boundary conditions at the two ends of the string. To see the implication of the single-particle dynamics, we also implement the infalling boundary condition to the one branch of our solutions to obtain the exact retarded Green's function as in \cite{Son_09,Lee_19}. We interpret this Green's function as a response function for a single particle resulting from integrating out the degrees of freedom of both the field and the other particle. In what follows, the dispersion relation of the particle of the entangled pair can presumably be decoded from the exact retarded Green's function through the Langevin equations.
In future work, we will derive these Langevin equations by taking the variation of the effective action of the particles in this paper. We may use the detailed dynamics from these equations to study, for instance, the time evolution of the entanglement entropy between two particles as in \cite{Lee_19}. In this paper, we focus on the exact results of the force-force correlators and the retarded Green's function with their implications for particle dynamics.  Based on the interpretations above, we find that the causality in the string worldsheet plays a crucial role in determining the nature of the dispersion relation of the particles and also the feature of the induced mutual interaction between them. In the case of an accelerating string with the causally disconnected two ends, the induced effect from the field gives the dissipative dynamics for the particles with no dependence on the distance between them. As a comparison, in the case of a static string, the particle dynamics is dissipative at the early enough times when two ends are not in causal contact with each other.
The induced dispersion relation of each particle becomes non-dissipative when two ends are causally connected at late times.  Moreover, in both cases, at late times, the induced mutual coupling between two ends, which comes from the field correlator or the force-force correlator, follows a power law to vanish. It will also be interesting to see whether this different time dependence of the field correlators and causality play any role in the black hole information problem proposed in \cite{Maldacena_01}.

The remaining part of the paper is organized as follows. Section \ref{accelerating} reviews the exact solution of the accelerating string and then solve the linearized equation for the perturbations in this worldsheet background. We obtain the all-time two-point correlators of the fields that the quark and anti-quark pair couple to, in contrast to the earlier works where only the late-time behavior of the correlators was found \cite{Xiao_08}.  As a comparison, we also find the exact solution of the linearized equation of the string worldsheet in the case when its two ends are static in section \ref{static string}. We then conclude and comment in section \ref{conclude}.

\section{fluctuations of transverse modes in the accelerating string }\label{accelerating}

We first review the fundamental equations and their exact solutions for the accelerating string in $AdS_{d+1}$ space \cite{Xiao_08}. Then we exactly solve the linearized equation for the perturbations in these string backgrounds and give a prescription to obtain the generating functional for the boundary fields coupled to the string endpoints.

\subsection{The exact solutions of the string background and perturbations}
\begin{figure}[h]
\centering
\includegraphics[scale=0.7]
{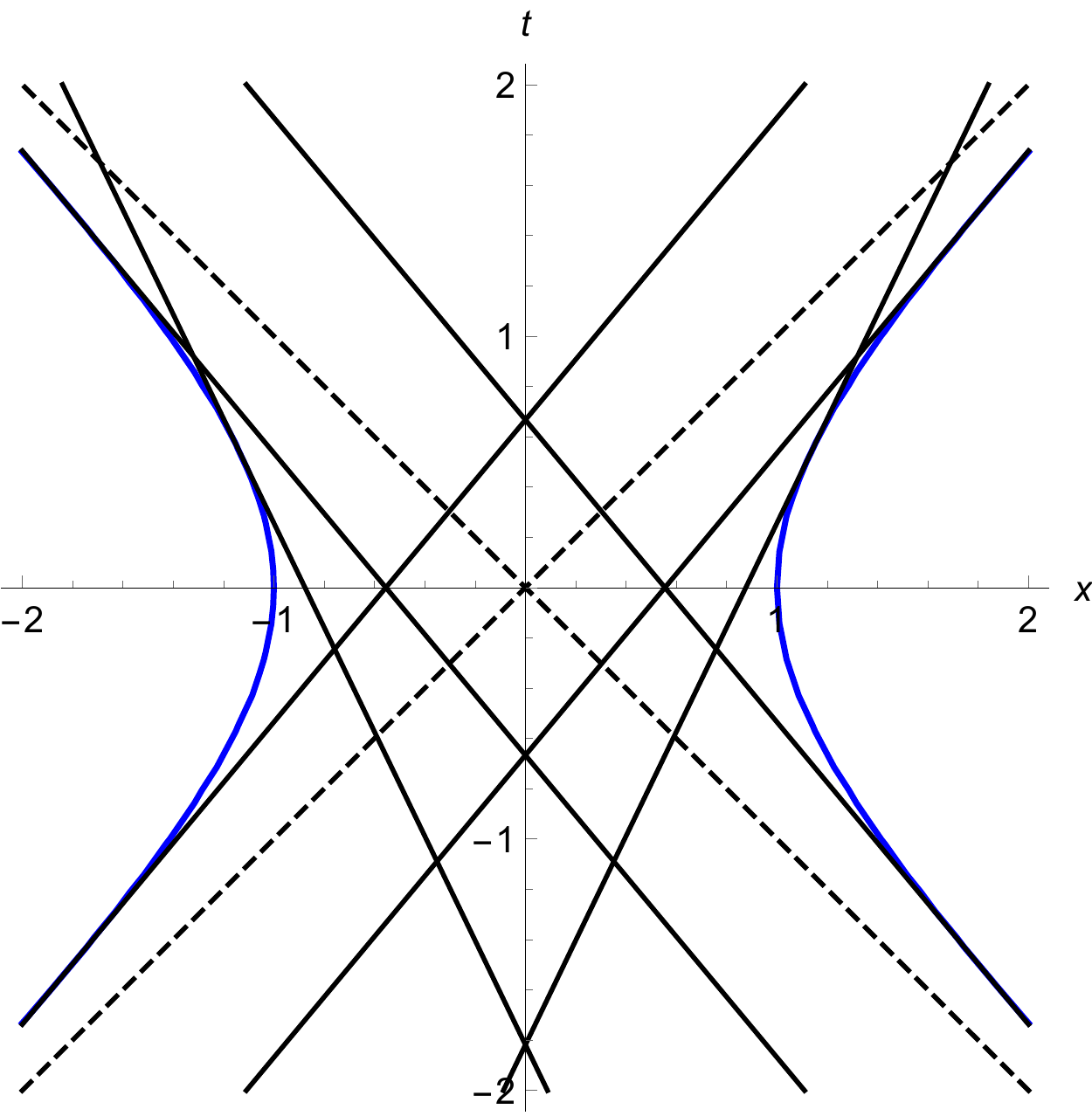} \caption{Few sample light rays (solid black lines) in the worldsheet(\ref{induce}). Here $x$ and $t$ are in unit of $b$. The hyperbolic curves (blue) are trajectories of quark and anti-quark. Two quarks are causally disconnected with the horizons at $t=\pm x$ (dash lines).}
\label{fig:world-sheet-causal-structure} 
\end{figure}

We consider the $AdS_{d+1}$ metric in the Poincare coordinates given by
  \be
  \label{ads}
  ds^2=\frac{R^2}{z^2}(-dt^2+dz^2+dx^2+\sum_{i=1}^{d-2}dy_i^2)\,  ,
  \ee
where $R$ is the curvature radius. The string in this background is described by the Nambu-Goto action
\begin{equation}
S=-T_0 \int d\tau d\sigma \sqrt{- h} \,,
\end{equation}
where $T_0$ is the string tension and $h=\text{det} \, h_{ab}$ is the determinant of the induced metric. We choose a static gauge $(\tau,\sigma)=(t,z)$ and the embedding of the string as $X^{\mu}(t,z)=(t,z,x(t,z),0,\cdots,0)$ with $\mu=1, 2, 3,\cdots, d$.
Then, $\sqrt{-h}=\frac{R^2}{z^2} \sqrt{1-\dot x^2+x'^2}$.
The classical equation of motion is given by
\be
\label{eom}
\frac{\partial}{\partial t} \Big(\frac{\dot x}{z^2 \sqrt{1-\dot x^2+x'^2}}\Big)-\frac{\partial}{\partial z} \Big(\frac{ x'}{z^2 \sqrt{1-\dot x^2+x'^2}}\Big)=0 \,,
\ee
where $'=\frac{\partial}{\partial z}$ and $\cdot=\frac{\partial}{\partial t}$.
This equation has an exact solution \cite{Xiao_08},
  \be
  \label{ws}
  x_b(t,z)=\pm\sqrt{t^2+b^2-z^2} \,,
  \ee
where $b$ is a real constant. In the solution, the trajectory of the end-points at $z=0$ describes the motion of two particles along the $x$-direction with uniform deceleration $\frac{1}{b}$. They head toward each other from $x=\pm\infty$ when $t=-\infty$, subsequently stop at $x=\pm b$ when $t=0$ and then turn around in the opposite directions, moving away from each other. We label the position of the particle moving in the positive (negative) $x$ region by $q_R (q_L)$, respectively.
The induced metric on the worldsheet with the embedding coordinates is
   \be
   \label{induce}
    ds_{ws}^2=\frac{R^2}{(t^2+b^2-x^2)^2}((x^2-b^2)dt^2-2tx\, dt\, dx+(t^2+b^2)dx^2) \,.
   \ee
As noticed in \cite{Jensen_13}, this metric has the same casual structure as that of the eternal two-sided $AdS$ back hole (see Fig.~\ref{fig:world-sheet-causal-structure}) and has the bifurcating horizons located at $t=\pm x$, the time-like boundaries at the particle trajectories $x^2-t^2=b^2$, and the space-like ``singularity'' at $x^2-t^2\rightarrow -\infty$. However, this space-like singularity is not a physical singularity and just reflects the coordinate singularity in the background AdS as $z\rightarrow\infty$ to approach the Poincare Killing horizon. Due to the analog between the EPR pair and the wormhole geometry \cite{ER=EPR}, it was proposed in \cite{Jensen_13, Sonner_13} that this metric (\ref{induce}) gives the gravity description of the entangled quark and anti-quark pair.

We now consider the fluctuations of a string in the direction transverse to the $x$ coordinate, say $y_1$, with the embedding in (\ref{ads}) as $X^{\mu}=(t,z,x_b(t,z),q(t,z),0,\cdots, 0)$ where $q$ is small as compared to $x_b$.
The quadratic action for $q$ is obtained as
  \be
  S_q=-R^2T_0\int dt\, dz \frac{1}{z^2}\left(\frac{q'^2-\dot q^2-\dot x_b^2 q'^2-x'^2_b \dot q^2+2\dot x_b x'_b \dot q q'}{2\sqrt{1-\dot x_b^2+x_b'^2}}\right) \, .
  \ee
 As seen in the action, in this coordinate system, there is no separable solution in the variables $t$ and $z$.  However, in the comoving frame of one of the accelerating particles, say the $R$ branch in (\ref{ws}),  defined as,
  \bea
  \label{coord}
  &&x=\sqrt{b^2-r^2}e^{\frac{\alpha}{b}}\cosh\frac{\tau}{b} \,,\\
  &&t=\sqrt{b^2-r^2}e^{\frac{\alpha}{b}}\sinh\frac{\tau}{b} \,,\\
  &&z=re^{\frac{\alpha}{b}} \,,
  \eea
with $0<r<b$, the metric of (\ref{ads}) becomes
  \be
  \label{BHmetric}
  ds^2 =\frac{R^2}{r^2}\left(-\Big(1-\frac{r^2}{b^2}\Big)d\tau^2+\frac{dr^2}{1-\frac{r^2}{b^2}}+d\alpha^2+e^{-\frac{2\alpha}{b}}\sum_{i=1}^{d-2}dy_i^2\right) \,.
  \ee
As in \cite{Xiao_08}, this metric has the event horizon at $r=b$ with the Hawking temperature $T=\frac{1}{2\pi b}$, which shows no causal connection between the two particles. However, on the worldsheet (\ref{induce}), it can be seen that a wormhole connects them. We also expect that in the particle rest frame, the particles will experience quantum fluctuations at finite temperature $T$ to be verified later.
Note that this temperature $T$ can also be interpreted as the Unruh temperature that an accelerating particle observes.
In this coordinate system, the background worldsheet solution in (\ref{ws}) is
  \be
  \alpha(\tau,r)=0\, .
  \ee
And the quadratic action for the perturbations in the $y_1$ direction on the worldsheet becomes
  \be
  S_{qc}=-\frac{R^2T_0}{2}\int d\tau dr \frac{1}{r^2}\left(\Big(1-\frac{r^2}{b^2}\Big)q'^2-\frac{\dot q^2}{1-\frac{r^2}{b^2}}\right)  \,
  \ee
with $'=\frac{\partial}{\partial r}$ and $\cdot=\frac{\partial}{\partial \tau}$.
In terms of the Fourier modes
  \be
  q(\tau,z)=\int\frac{d\omega}{2\pi} \, y_{\omega}(z)\, e^{-i\omega\tau} \,,
  \ee
  the equation of motion reads
  \be
  \label{eom1}
  y''_{\omega}-\frac{2}{z (1-\frac{z^2}{b^2})}y'_{\omega}+\frac{\omega^2 }{(1-\frac{z^2}{b^2})^2}y_{\omega}=0 \, .
  \ee
On the worldsheet, the comoving coordinate $r$ is equal to the static coordinate $z$, and now $'=\frac{\partial}{\partial z}$.
To make the action finite, we introduce a cutoff scale by putting the boundary brane at the finite $z=z_m$. Then, with the normalization condition $y_{\omega}(z_m)=1$, the equation (\ref{eom1}) has two independent solutions
  \be
  \label{sol}
  Y_{\omega}(z)=\frac{(1-i\omega z)e^{i\omega b \tanh^{-1}\frac{z}{b}}}{(1-i\omega z_m)e^{i\omega b \tanh^{-1}\frac{z_m}{b}}}~~~\mbox{and}~~~Y_{\omega}^*(z) \, ,
  \ee
where $Y_{\omega}(z)$ ($Y^*_{\omega}(z)$) is the incoming (outgoing) wave near the horizon $z=b$. We would like to stress that these are the exact solutions, whereas in  \cite{Xiao_08} only the approximate solutions in the small-$\omega$ expansion are given.

\subsection{Single-particle dynamics}

Using the holographic prescription, the retarded Green's function of the fields coupled to one end of the string is
  \bea
  \label{GR}
  G_R(\omega)=\frac{R^2T_0}{b^2}\left(\frac{b^2}{z_m^2}-1\right) Y'_{\omega}(z_m)Y^*_{\omega}(z_m)&=&\frac{R^2T_0}{z_m b^2}\frac{1}{1+\omega^2z_m^2}\left(\omega^2(b^2-z_m^2)+i(z_m\omega+b^2z_m\omega^3)\right)  \nonumber\\
  &=& i\gamma_T \, \omega + M_T \omega^2 + \mathcal{O}(\omega^3) \,,
  \eea
which is analytic in the upper half complex $\omega$-plane. Its small $\omega$ limit agrees with the results in \cite{Xiao_08} and also our previous papers \cite{Lee_13}, giving the effects to the particle of the temperature-dependent damping term $\gamma_T=R^2 T_0/b^2$ and the effective mass term $M_T=M_0+\Delta M_T$.
$M_T$ consists of the large temperature-independent mass $M_0=R^2 T_0/z_m$ and the small temperature-dependent mass correction $\Delta M_T=-R^2 T_0 z_m/b^2$  in the limit of $z_m/b \rightarrow 0$.  This picture of the particle dynamics is captured in the Langevin equation \cite{Son_09,Lee_19},
  \be
  \int_0^{\tau}G_R(\tau-\tau')q(\tau')d\tau'=\eta(\tau)
  \ee
where $q(\tau)$ is the position of the particle and $\eta(\tau)$ is the noise force with $\langle \eta(\tau)\eta(\tau')\rangle$ given by the Hadamard function of the field, $G_H(\tau-\tau')$. This can also be used to obtain the particle correlator, which in frequency space is given by
    \be
    \label{qq1}
    \langle \tilde{q}(\omega)\tilde{q}(-\omega)\rangle=\frac{G_H(\omega)}{G_R(\omega)G_R(-\omega)} \,.
    \ee
This equation incorporates not only the dissipation effect given by the retarded Green's function $G_R$ of the field in (\ref{GR}) but also the force-force correlation function in terms of the Hadamard function that satisfies the fluctuation-dissipation theorem, namely $G_H (\omega)= \coth (\omega/T)\, {\rm {Im}} G_R (\omega)$ (this relation was shown to be true in very general holographic setups \cite{Dimitrio_18}). To estimate the late time behavior of the particle correlator, we consider the retarded Green's function in the small $\omega$ expansion in (\ref{GR}). Through the fluctuation-dissipation theorem, the Hadamard function  $G_H \simeq T\gamma_T$. So we have $\langle \tilde{q}(\omega)\tilde{q}(-\omega)\rangle\propto \omega^{-2}$. Thus, to obtain the finite particle correlator, we need to introduce an infrared cutoff, for example, by adding a potential trap $\Omega^2q^2(\tau)$ to the end of the string\footnote[1]{Another way to introduce the cutoff is by deforming the integral to the complex $\omega$-plane, where the double pole contribution at $\omega=0$ also gives the correlator linear in $\tau$. This prescription can be naturally incorporated in defining different type of Green's functions by the contour integrals. Here, the potential trap cut-off is more natural as we want to make the saturated entanglement entropy finite as in \cite{Lee_19}.}, and this gives in the small $\omega$ limit,
  \be
  \langle q(\tau)q(0)\rangle \propto \int_{-\infty}^{\infty}\frac{e^{-i\omega\tau}}{\omega^2+\Omega^2}d\omega=\pi\Omega^{-1}e^{-\Omega \tau} \,.
  \ee
When taking away the cutoff, $\Omega\rightarrow 0$, we have the cutoff independent term that grows linearly in $\tau$. This signals the late-time growth of uncertainty in the particle position for observers moving together with the particle. Notice that here $\omega$ is the frequency corresponding to the proper time for the particle. For the particle correlator in the inertial time, $t$, we can see from (\ref{time}) and the fact that in the late time $\tau=b\ln(t/b)$,
  \be
  \label{qq}
  \langle q(t)q(0)\rangle_{cutoff-independent} \propto t^{-1}\ln(t/b) \,.
  \ee
In \cite{Lee_19}, we have used this particle correlator, in the setup of strings with only one end on the boundary, to calculate the entanglement entropy between particles and environment fields in the late times, and in that case the divergent piece in $\langle q(t)q(0)\rangle$ is absorbed into the renormalized saturated entropy. Here, we are more interested in the dynamics of two entangled particles in the current paper, but we will leave the detailed study for future works. For the moment, we can qualitatively see the rate for information exchange between a single particle and the field to which it couples.

Furthermore, as far as we know, via a bilinear coupling between a particle and environmental fields, the temperature-dependent damping term can only be achieved from the holographic approach, whereas in the free field theory, the same type of the coupling leads to the state-independent back reaction from the field to the particle \cite{JT_1}. It is also worth mentioning that when two ends of the string undergo the acceleration, they are never in causal contact during their journeys. In this case, the energy of the particle transferred to the field does not have a chance to come back, which gives the dissipative dynamics of the particles. Moreover, due to the causality consideration, the retarded Green's function of the particle has no dependence on the distance between two particles. In contrast, in the static string case, two ends are initially causally disconnected and then become connected at the time scale of the distance between two ends. It thus induces a somewhat different dispersion relation of the particle after two ends reach the causal contact.

\subsection{The generating functional of fields}

To obtain the force-force correlations between two ends of the string, which encode the information of the wormhole, we need to find the generating functional for the field that couples to them. For this purpose, we transform the solution in (\ref{sol}) back to the coordinates in (\ref{ads}).  The $R(L)$ branch corresponds to the $+(-)$ background solution in (\ref{ws}). Also, notice that the $R$ and $L$ branches have different relations between the proper time $\tau$ and the coordinate time $t$.
 Thus, from (\ref{coord})  $\tau_{R/L}=\mp\tanh^{-1}\frac{t}{\sqrt{b^2+t^2-z^2}}$.  Then we can write the general solutions as
  \be
  q^{R/L}(t,z)=\int \frac{d\omega}{2\pi}e^{\mp i\omega b \tanh^{-1}\frac{t}{\sqrt{b^2+t^2-z^2}}}\left(f^{R/L}(\omega)Y_{\omega}(z)+g^{R/L}(\omega)Y^*_{\omega}(z)\right)\, .
  \ee
The boundary values of $q^{R/L}(t,z)$ are interpreted holographically as the sources for the quark and anti-quark, $q^{R/L}(t,z_m)\equiv q_0^{R/L}(t)$. And their ``Fourier transform'' is defined as
  \be \label{FT}
  q_0^{R/L}(t)=\int\frac{d\omega}{2\pi}e^{\mp i\omega u(t)}\tilde{q}_0^{R/L}(\omega) \,,
  \ee
where $u(t)\equiv b\tanh^{-1}\frac{t}{\sqrt{b^2+t^2-z_m^2}}$. Thus we have two boundary conditions at $z=z_m$ as
   \be
   \label{bc1}
   f^{R/L}(\omega)+g^{R/L}(\omega)=\tilde{q}_0^{R/L}(\omega)\, .
   \ee
The other two boundary conditions for $q^{R/L}(t,z)$ are imposed at the horizon $z=b$.
As in \cite{Son_03}, the analyticity properties of the boundary conditions for the mode functions cross the horizon of a maximally extended black hole gives the Schwinger-Keldysh correlators for a single particle. In our previous work \cite{Lee_15}, we also obtained the same set of correlators by imposing the constraints from the unitarity and periodicity of the finite temperature boundary correlators. In the setting of this paper, the maximally extended black hole does not describe the double copy of a single particle action giving the Schwinger-Keldysh correlators, but the action of the two coupled particles. To achieve it, we
will modify the definition of the Kruskal coordinates in \cite{Son_03}. We will also give the prescription from the field theory constraints similar to the ones in \cite{Lee_15}.

Near the horizon in (\ref{BHmetric}), the mode functions can be simplified when  the Kruskal coordinates, say in the $R$ branch, are introduced  as
  \be
  V=-e^{-\frac{\tau+r^*}{b}},~~~U=e^{\frac{\tau-r^*}{b}} \, ,
  \ee
where $r^*=b\tanh^{-1}\frac{r}{b}$. Then  the mode function can be written as
  \be
  q^R\simeq f^R(\omega)(U)^{-ib\omega}+g^R(\omega)(-V)^{ib\omega} \, .
  \ee
Notice that $UV=\frac{r-b}{r+b}$. Therefore, we can even define the mode functions inside the horizon when $UV>0$. A similar definition of the mode functions for the $L$ branch is adopted, where now  $r^*=-b\tanh^{-1}\frac{r}{b}$. So, the value of $r^*$ is negative in the $L$ branch and becomes positive in the $R$ branch. In the $L$ branch,
    \be
  V=e^{\frac{\tau+r^*}{b}},~~~U=-e^{-\frac{\tau-r^*}{b}} \,.
  \ee
Note that $\tau$ in the $L$ branch has an opposite sign relative to the $R$ branch.
Then the mode function in the $L$ branch can be written down as
  \be
  q^L\simeq f^L(\omega)(V)^{ib\omega}+g^L(\omega)(-U)^{-ib\omega}\, .
  \ee
To match $q^R$ with $q^L$, we require them to be analytic on the real $U$ and $V$ axes and the lower half of complex $U$ and $V$ planes as in \cite{Son_03}. Thus, the matching conditions are given by
\be
  \label{bc2}
  f^{R}(\omega)=e^{-\pi b\omega}g^{L}(\omega),~~~g^{R}(\omega)=e^{\pi b\omega}f^{L}(\omega) \, ,
  \ee
and together with (\ref{bc1}), they uniquely fix the bulk mode functions $q^{R/L}(t,z)$ with
    \bea
  &&f^R(\tilde{\omega})=\frac{\tilde{q}^R_0(\omega)}{1-e^{2b\pi\omega}}-\frac{e^{b\pi\omega}\tilde{q}^L_0(\omega)}{1-e^{2b\pi\omega}} \,,\\
  &&g^R(\tilde{\omega})=\frac{e^{b\pi\omega}\tilde{q}^L_0(\omega)}{1-e^{2b\pi\omega}}-\frac{e^{2b\pi\omega}\tilde{q}^R_0(\omega)}{1-e^{2b\pi\omega}} \,,\\
  &&f^L(\tilde{\omega})=\frac{\tilde{q}^L_0(\omega)}{1-e^{2b\pi\omega}}-\frac{e^{b\pi\omega}\tilde{q}^R_0(\omega)}{1-e^{2b\pi\omega}} \,,\\
  &&g^L(\tilde{\omega})=\frac{e^{b\pi\omega}\tilde{q}^R_0(\omega)}{1-e^{2b\pi\omega}}-\frac{e^{2b\pi\omega}\tilde{q}^L_0(\omega)}{1-e^{2b\pi\omega}} \,.
  \eea

The bulk on-shell action for this solution, which contains only the boundary terms, is identified as a generating functional for the boundary field correlators. Leaving out the contact terms that should be related to the renormalization of the particle self-energy, we have
 \bea
  S[\tilde{q}^R_0,\tilde{q}^L_0]&=&S_q(\tilde q_0^R)+S_q(\tilde q_0^L)\\
  &=&\frac{1}{2}\int \frac{d\omega}{2\pi}\{ \tilde{q}^R_0(\omega)[\mbox{Re} G_R(\omega)+\frac{e^{\frac{\omega}{T}}+1}{e^{\frac{\omega}{T}}-1}i\mbox{Im}G_R(\omega)]\tilde{q}^R_0(-\omega)\\
  &&+\tilde{q}^R_0(\omega)[\frac{-2ie^{\frac{\omega}{2T}}}{e^{\frac{\omega}{T}}-1}\mbox{Im}G_R(\omega)]\tilde{q}^L_0(-\omega)\\
  &&+\tilde{q}^L_0(\omega)[\frac{-2ie^{\frac{\omega}{2T}}}{e^{\frac{\omega}{T}}-1}\mbox{Im}G_R(\omega)]\tilde{q}^R_0(-\omega)\\
   &&+\tilde{q}^L_0(\omega)[\mbox{Re} G_R(\omega)+\frac{e^{\frac{\omega}{T}}+1}{e^{\frac{\omega}{T}}-1}i\mbox{Im}G_R(\omega)]\tilde{q}^L_0(-\omega)\}
  \eea
with $G_R(\omega)$ given in (\ref{GR}).  Then we can read off the correlators $G^{ij}(\omega)=\frac{\delta}{\delta\tilde{q}^i_0}\frac{\delta}{\delta\tilde{q}^j_0}S[\tilde{q}^R_0,\tilde{q}^L_0]$ with $i,j=R,L$, and the real-time correlators:
   \be
   \label{time}
   G^{ij}(t,t')=\frac{(-1)^i b}{\sqrt{b^2+t^2-z_m^2}}\frac{(-1)^j b}{\sqrt{b^2+t'^2-z_m^2}}\int\frac{d\omega}{2\pi}G^{ij}(\omega)e^{-i\omega(\tau^i(t)-\tau^j(t'))} \,.
   \ee
Here, $\tau^R(t)=u(t)$ and $(-1)^R=1$, while $\tau^L(t)=-u(t)$ and $ (-1)^L=-1$.

From the field theory point of view, we consider the particles $ \tilde{q}_0^{R/L}$ coupled to the common quantum field at finite temperature through a bilinear coupling with the same coupling strength.
Integrating out the degrees of freedom of the field leads to the effective action of the particles.
Thus, the generating function of the field requires to be symmetric under the exchange of $R$ and $L$, leading to $G^{RL}=G^{LR}$ and $G^{RR}=G^{LL}$. Also, the Green's functions $G^{RR}$ and $G^{LL}$ should have the form of the finite temperature time-ordered Green's functions at temperature $T=\frac{1}{2\pi b}$. The above requirements turn out to be equivalent to (\ref{bc2}) from the bulk analyticity condition.

\begin{figure}[h]
\centering
\includegraphics[scale=0.7]
{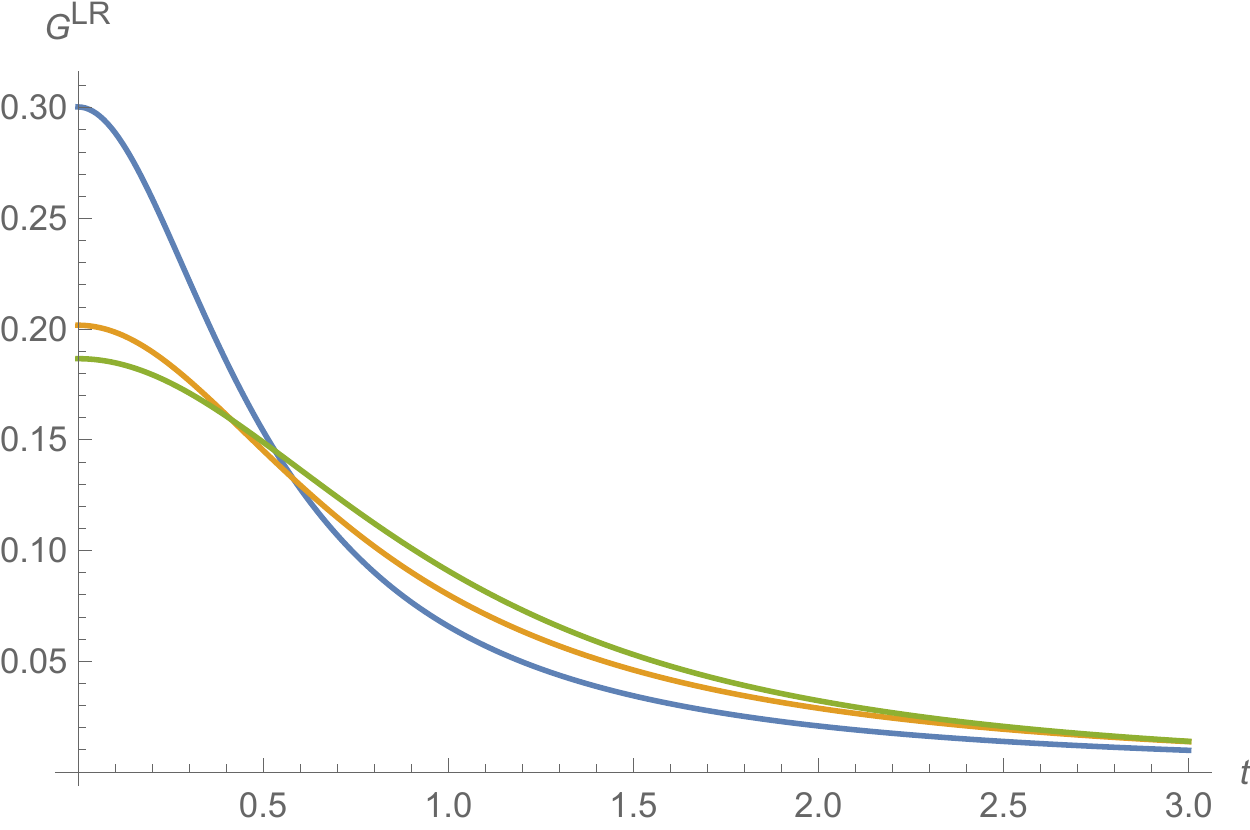} \caption{The time evolution of the $G^{LR}(t,0)=G^{RL}(0,t)$ correlator. $t$ is in the unit of $b$ and $G^{LR}$ is in the unit of $\frac{4R^2T_0i}{b^4\sqrt{1-\frac{z_m^2}{b^2}}}$. The curves correspond to the parameters $\frac{z_m}{b}=0.9$(blue), $0.7$(orange) and $0.1$(green), respectively.}
\label{fig:plot_G_LR} 
\end{figure}

Notice that $G^{RR}(\omega)=G^{LL}(\omega)$ has the same form as the finite-temperature time-order correlators. Even though we have not studied the coupled Langevin equations, the correlation between two entangled particles is, in principle, encoded in the cross correlators, $G^{LR}(\omega)$ and $G^{RL}(\omega)$, just like in the case of the single-particle dynamics (\ref{qq1}). $G^{LR}(t,0)=G^{RL}(0,t)$ is plotted numerically in Fig.~\ref{fig:plot_G_LR}.

We turn to the early-time and late-time analytic behaviors of the $G^{LR}(t,0)=G^{RL}(0,t)$ correlators.
Note that the retarded Green's function, $G_R(\omega)$ in (\ref{GR}), has the pole at $\omega=- \frac{i}{z_m}$, and the Boltzmann factor gives the poles at $\omega= \frac{in}{b}$ with non-zero integers. We take $b\gg z_m$, where $b$ plays a role of an IR cut-off and $z_m$ a UV cut-off. Thus, in the limit $t\ll z_m$ with the proper time $u(t)\simeq t$, the time evolution of $G^{LR}(t,0)=G^{RL}(t,0)$  is dominated by the pole at $\omega=-\frac{i}{z_m}$ giving
  \be
  \label{G1}
  G^{LR}(t,0)= G^{RL}(t,0)\propto e^{-\frac{t}{z_m}}\simeq 1-\frac{t}{z_m}+\cdots \,.
  \ee
As for $t\gg b$ with $u(t)\simeq b\ln(t/b)$, the pole at $\omega=\frac{-i}{b}$ dominates the correlators $G^{ij}(t,0)$ resulting in the power-law decay
\be
\label{G2}
G^{LR}(t,0)= G^{RL}(t,0)\propto\left(\frac{t}{b}\right)^{-2} \, .
\ee

This can be compared with the correlators of the free scalar fields on the spacetime points of two observers following the trajectories, say along the $x$-direction with the uniform acceleration $1/b$. The worldlines of two observers are specified by $x_L^\mu=( b \sinh \frac{\tau_L}{b}, - b \cosh \frac{\tau_L}{b}, 0, 0)$ and $x_R^\mu=( b \sinh \frac{\tau_R}{b}, b \cosh \frac{\tau_R}{b}, 0, 0)$. In particular, the correlator of the free scalar field at the spacetime point $x_L^\mu$ and $x_R^\mu$ is given by
\be
G(x_L^\mu,x_R^\mu)=\frac{1}{4\pi^2} \frac{1}{\vert \vec x_R-\vec x_L\vert^2-\vert  x_R^0- x_L^0 \vert^2}\, .
\ee
Thus, when $t\ll b$ giving the proper time $\tau \simeq t$,
\be
  G(t,0)\propto \left( 1- \frac{t^2}{2 b^2} +...\right) \,,
\ee
whereas for $t\gg b$ giving $\tau \simeq  b\ln(t/b)$,
\be
G (t,0)\propto\left(\frac{t}{b}\right)^{-1} \,
\ee
instead. By comparing with (\ref{G2}), we see that for strongly coupled fields the correlation between  fields at two spacelike-separate points dies out more quickly than the one for free fields, and this potentially induces the interaction between two spacelike-separate particles and gives a shorter time scale for the particles to sustain quantum entanglement between them.

\section{fluctuations of transverse modes in the static string}\label{static string}
\begin{figure}[h]
\centering
\includegraphics[scale=0.6]
{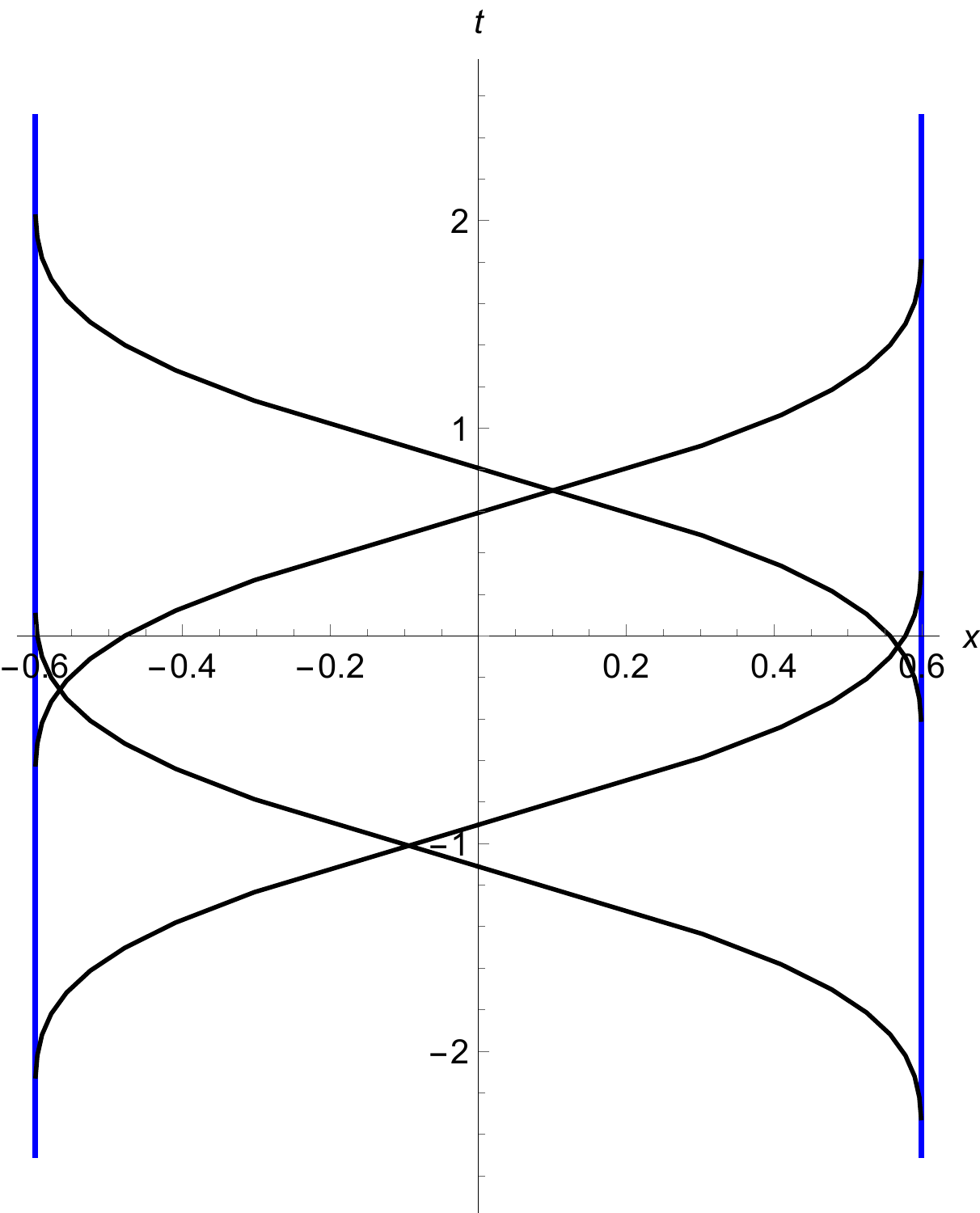} \caption{Few sample light rays (black curves) in the worldsheet(\ref{zeroT}). Here $x$ and $t$ are in the unit of $z_0$. Two vertical lines (blue) are trajectories of quark and anti-quark. Light rays emitted from one quark reach the other in some finite times. So two quarks are causally connected.}
\label{fig:world-sheet-causal-structure-static} 
\end{figure}

This section considers the static string in the same $AdS_{d+1}$ background in (\ref{ads}). We also solve the linearized equation for the perturbations in this string background exactly and obtain the generating functional for the field that couples to the two ends of the string. In the same gauge choice as in (\ref{eom}), where the coordinate $x$ is independent of time, the classical equation of motion for the static string with its two ends at the boundary $z=0$ has a solution with the two-branch joint at $z=z_0$ \cite{Maldacena:1998im},
    \be
    \label{zeroT}
    x_0(z)=\pm\int_{z_0}^{z}\sqrt{\frac{y^4}{z_0^4-y^4}}dy \, .
    \ee
The induced metric on this worldsheet parametrized by $t$ and $x$ has the causal structure as in Fig.~\ref{fig:world-sheet-causal-structure-static}. As compared with the case of the accelerating string (in Fig.~\ref{fig:world-sheet-causal-structure}), we see that two ends are in causal contact in this case.

Similar to the accelerating string case, we consider the linearized perturbations in the direction transverse to the worldsheet, with the embedding in (\ref{ads}) as $X^{\mu}=(t,z,x_0(z),p(t,z),0,0...)$ where $p$ is small as compared to $x_0$. We define the Fourier transform of the transverse modes as
\be
p(t,z)=\int\frac{d\omega}{2\pi}h_{\omega}(z)e^{-i\omega t} \, ,
\ee
 which has the linearized equation of motion on both $R$ and $L$ branches as,
  \be
  \label{static}
  h''_{\omega}-\frac{2}{\rho(1-\rho^4)}h'_{\omega}+\frac{\tilde{\omega}^2}{1-\rho^4}h_{\omega}=0
\,,
  \ee
with the dimensionless variables $\rho=\frac{z}{z_0}$ and $\tilde{\omega}=z_0\omega$. There are two exact independent solutions, which are normalized in the boundary at $\rho=\rho_m\equiv\frac{z_m}{z_0}$,
 \be
 F_{\tilde{\omega}}(\rho)=e^{A_{\tilde{\omega}}(\rho)-A_{\tilde{\omega}}(\rho_m)}\, , \quad\quad H_{\tilde{\omega}}(\rho)=e^{B_{\tilde{\omega}}(\rho)-B_{\tilde{\omega}}(\rho_m)} \,
 \ee
where
  \bea
  &&A_{\tilde{\omega}}(\rho)=\frac12\ln(1+\tilde{\omega}^2\rho^2)+\frac{\sqrt{1-\tilde{\omega}^4}}{\tilde{\omega}}(E_0(\rho)-E_{\tilde{\omega}}(\rho))\, ,\\
  &&B_{\tilde{\omega}}(\rho)=\frac12\ln(1+\tilde{\omega}^2\rho^2)-\frac{\sqrt{1-\tilde{\omega}^4}}{\tilde{\omega}}(E_0(\rho)-E_{\tilde{\omega}}(\rho)) \,,
  \eea
with
  \be
  E_0(\rho)=\int_{0}^{\rho}\frac{dy}{\sqrt{1-y^4}}\, ,\quad\quad E_{\tilde{\omega}}(\rho)=\int_{0}^{\rho}\frac{dy}{(1+\tilde{\omega}^2y^2)\sqrt{1-y^4}} \, .
  \ee
Then the general solutions of the perturbations on the $R$ and $L$ branches are written as
  \be
  p^{R/L}(t,\rho)=\int\frac{d\omega}{2\pi}\left(C_1^{R/L}(\tilde{\omega}) F_{\tilde{\omega}}(\rho)+C_2^{R/L}(\tilde{\omega}) H_{\tilde{\omega}}(\rho)\right)e^{-i\omega t}\, .
  \ee
The boundary values of the perturbations are interpreted as the sources for the quark and anti-quark pair,
 \be
 p^{R/L}(t,\rho_m)=p^{R/L}_0(t)=\int\frac{d\omega}{2\pi} \, \tilde{p}^{R/L}_0(\omega) \, e^{-i\omega t} \, .
 \ee
Thus, we have the boundary conditions
   \be
   \label{bdy1}
   C_1^{R/L}(\tilde{\omega})+C_2^{R/L}(\tilde{\omega})=\tilde{p}^{R/L}_0(\tilde{\omega})
\,.   \ee
The other two boundary conditions are imposed at the tip of the string ($\rho=1$). We require the perturbations to be smooth across the tip, namely $p^{R}(t,1)=p^{L}(t,1)$ and $\partial_{\rho}p^{R}(t,1)=-\partial_{\rho}p^{L}(t,1)$, which lead to
  \bea
  &&C_1^{R}(\tilde{\omega})F_{\tilde{\omega}}(1)+C_2^{R}(\tilde{\omega})H_{\tilde{\omega}}(1)=C_1^{L}(\tilde{\omega})F_{\tilde{\omega}}(1)+C_2^{L}(\tilde{\omega})H_{\tilde{\omega}}(1) \,, \\
  &&C_1^{R}(\tilde{\omega})F_{\tilde{\omega}}(1)-C_2^{R}(\tilde{\omega})H_{\tilde{\omega}}(1)=-C_1^{L}(\tilde{\omega})F_{\tilde{\omega}}(1)+C_2^{L}(\tilde{\omega})H_{\tilde{\omega}}(1) \label{antibc} \label{bdy2} \,.
  \eea
Note that to obtain (\ref{antibc}), we have factored out a common divergence factor.
Together with (\ref{bdy1}), the unique solution can be obtained as
  \bea
  &&C^R_1(\tilde{\omega})=\frac{\tilde{p}^R_0(\tilde{\omega})}{1-e^{2\kappa(\tilde{\omega})}}-\frac{e^{\kappa(\tilde{\omega})}\tilde{p}^L_0(\tilde{\omega})}{1-e^{2\kappa(\tilde{\omega})}} \,, \\
  &&C^R_2(\tilde{\omega})=\frac{e^{\kappa(\tilde{\omega})}\tilde{p}^L_0(\tilde{\omega})}{1-e^{2\kappa(\tilde{\omega})}}-\frac{e^{2\kappa(\tilde{\omega})}\tilde{p}^R_0(\tilde{\omega})}{1-e^{2\kappa(\tilde{\omega})}} \,, \\
  &&C^L_1(\tilde{\omega})=\frac{\tilde{p}^L_0(\tilde{\omega})}{1-e^{2\kappa(\tilde{\omega})}}-\frac{e^{\kappa(\tilde{\omega})}\tilde{p}^R_0(\tilde{\omega})}{1-e^{2\kappa(\tilde{\omega})}} \,, \\
  &&C^L_2(\tilde{\omega})=\frac{e^{\kappa(\tilde{\omega})}\tilde{p}^R_0(\tilde{\omega})}{1-e^{2\kappa(\tilde{\omega})}}-\frac{e^{2\kappa(\tilde{\omega})}\tilde{p}^L_0(\tilde{\omega})}{1-e^{2\kappa(\tilde{\omega})}} \,,
  \eea
where $\kappa(\tilde{\omega})=\frac{2\sqrt{1-\tilde{\omega}^4}}{\tilde{\omega}}(E_0(1)-E_{\tilde{\omega}}(1)-E_0(\rho_m)+E_{\tilde{\omega}}(\rho_m))$.
Moreover, the on-shell action, which is identified as the generating functional of the field,  becomes
\bea
  &&S[\tilde{p}^R_0,\tilde{p}^L_0]=-\frac{R^2T_0}{2z^3_0}\lim_{\rho\rightarrow \rho_m}\int dt\frac{\sqrt{1-\rho^4}}{\rho^2}\left(p^{R}(t,\rho)\partial_{\rho}p^{R}(t,\rho)+p^{L}(t,\rho)\partial_{\rho}p^{L}(t,\rho)\right)\\
  &&=-\frac{R^2T_0}{2z^3_0} \frac{\sqrt{1-\rho_m^4}}{\rho_m^2}\int \frac{d\omega}{2\pi}\Big\{ \tilde{p}_0^R(\tilde{\omega})\left[\frac{1}{1-e^{2\kappa(\tilde{\omega})}}(A'_{\tilde{\omega}}(\rho_m)-e^{2\kappa(\tilde{\omega})}B'_{\tilde{\omega}}(\rho_m))\right]\tilde{p}_0^R(-\tilde{\omega})\\
  &&+\tilde{p}_0^R(\tilde{\omega})\left[\frac{e^{\kappa(\tilde{\omega})}}{1-e^{2\kappa(\tilde{\omega})}}(B'_{\tilde{\omega}}(\rho_m)-A'_{\tilde{\omega}}(\rho_m))\right]\tilde{p}_0^L(-\tilde{\omega})\\
  &&+\tilde{p}_0^L(\tilde{\omega})\left[\frac{e^{\kappa(\tilde{\omega})}}{1-e^{2\kappa(\tilde{\omega})}}(B'_{\tilde{\omega}}(\rho_m)-A'_{\tilde{\omega}}(\rho_m))\right]\tilde{p}_0^R(-\tilde{\omega})\\
   &&+\tilde{p}_0^L(\tilde{\omega})\left[\frac{1}{1-e^{2\kappa(\tilde{\omega})}}(A'_{\tilde{\omega}}(\rho_m)-e^{2\kappa(\tilde{\omega})}B'_{\tilde{\omega}}(\rho_m))\right]\tilde{p}_0^L(-\tilde{\omega})\Big\}\, .
  \eea
Based on the analyticity property of the mode functions,  we can identify the  retarded Green's function as
  \bea
  &&G^{(0)}_R(\omega)=-\frac{R^2T_0}{z^3_0}\frac{\sqrt{1-\rho_m^4}}{\rho_m^2} F'_{\tilde{\omega}}(\rho_m)F_{\tilde{\omega}}(\rho_m)\\
  &&=-\frac{R^2T_0}{z^3_0}\frac{\sqrt{1-\rho_m^4}}{\rho_m^2} A'_{\tilde{\omega}}(\rho_m)=-\frac{R^2T_0}{z^3_0}\frac{\sqrt{1-\rho_m^4}}{\rho_m^2}\,\, \frac{\tilde{\omega}\rho_m(\tilde{\omega}+\rho_m\sqrt{\frac{1-\tilde{\omega}^4}{1-\rho_m^4}})}{1+\tilde{\omega}^2\rho_m^2}\, ,
  \eea
 which is analytic in the upper half complex $\omega$ plane. Similarly, the advanced Green's function, which is  analytic in the lower half complex $\omega$ plane, can also be identified as
  \be
  G^{(0)}_A(\omega)=-\frac{R^2T_0}{z^3_0} \frac{\sqrt{1-\rho_m^4}}{\rho_m^2} B'_{\tilde{\omega}}(\rho_m)=-\frac{R^2T_0}{z^3_0}\frac{\sqrt{1-\rho_m^4}}{\rho_m^2} \,\, \frac{\tilde{\omega}\rho_m(\tilde{\omega}-\rho_m\sqrt{\frac{1-\tilde{\omega}^4}{1-\rho_m^4}})}{1+\tilde{\omega}^2\rho_m^2}\, .
  \ee
Thus, from the generating functional, we can read off the field  correlators arising from the field vacuum fluctuations as
 \bea
  &&G^{(0)RR}(\omega)=G^{(0)LL}(\omega)=\frac{1}{1-e^{2\kappa(\tilde{\omega})}}(G^{(0)}_R(\omega)-e^{2\kappa(\tilde{\omega})}G^{(0)}_A(\omega))\, ,\\
  &&G^{(0)RL}(\omega)=G^{(0)RL}(\omega)=\frac{e^{\kappa(\tilde{\omega})}}{1-e^{2\kappa(\tilde{\omega})}}(G^{(0)}_A(\omega)-G^{(0)}_R(\omega))\, .
 \eea

To see that these Green's functions are reasonable, we consider the limits of $\omega^{-1}\ll z_0$ and $z_0 \gg z_m$; they lead to $\tilde{\omega}\gg 1$ and $\rho_m\ll 1$, where the time-order correlators on the single particle are recovered in the limit of  $z_0 \rightarrow \infty$ as expected. With $\kappa(\tilde{\omega})\simeq 2i\gamma_1\tilde{\omega}$ and $\gamma_1\equiv E_0(1)=\frac{\sqrt{\pi}\Gamma(\frac54)}{\Gamma(\frac34)}$,  $G^{(0)}_R(\omega)= \big( G^{(0)}_A(\omega)\big)^*$ gives the Green's functions of the form
  \be
  G^{(0)RR}(\omega)=G^{(0)LL}(\omega)\simeq \mbox{Re}G^{(0)}_R(\omega)+\frac{1+e^{4i\gamma_1\tilde{\omega}}}{1-e^{4i\gamma_1\tilde{\omega}}}i\mbox{Im}G^{(0)}_R(\omega)\, .
  \ee
Note that this Green's function is not well defined since there are poles on the real $\omega$ values.  We thus employ the $i \varepsilon$ prescription. We replace $\omega$ by $\omega(1-i\varepsilon)$ and take $\varepsilon\rightarrow 0$ only at the end of the calculations.
Then, in the large $\tilde{\omega}$ limit, the Green's functions boil down to
   \be
   G^{(0)RR}(\omega)=G^{(0)LL}(\omega)\simeq \mbox{Re}G^{(0)}_R(\omega)+\mbox{sign}(\omega)i\mbox{Im}G^{(0)}_R(\omega)\, ,
   \ee
which is  precisely the Feynman Green's function at zero temperature.   In the limit of $\omega \gg z_0^{-1}$ with the relatively large distance between two particles but $\omega \ll z_m^{-1}$  for this small $\omega$ compared with the UV cutoff scale $z_m^{-1}$, the retarded Green's function is approximated by
  \bea
  G^{(0)}_R(\omega)& \simeq & -\frac{R^2T_0}{z_m^3}\frac{z_m^2\omega^2+iz_m^3\omega^3}{1+z_m^2\omega^2}\nonumber\\
  &\simeq & -M_0 \, \omega^2-i \gamma \, \omega^3 +\cdots \, .
  \eea
This gives the induced mass $M_0 \simeq R^2 T_0/z_m$, which has the same form as in the accelerating string case. The damping term of the $\omega^3$ dependence is a typical self-force effect for the Brownian motion with the cutoff-independent damping coefficient $\gamma \simeq R^2 T_0$, which also agrees with our previous results \cite{Lee_13}. Furthermore, this damping term can be compared with the one in the Abraham-Lorentz-Dirac equation of the charged particle with the self-force effects from the coupling to the electromagnetic field \cite{Guijosa_09,Guijosa_10,JT_1}. On top of that, as anticipated, the $z_0$ dependence drops off at this limit where two particles are far apart. Moreover, with the $i \epsilon$ prescription, the cross correlators $G^{(0)RL}(\omega)$ and $G^{(0)LR}(\omega)$ vanish in this limit.

In the other limits of $\tilde{\omega}\ll 1$ and $\rho_m\ll 1$, the wavelength of the perturbations is longer than the $z_0$ scale. The late-time behavior, when two ends are causally connected, is probed. In this case, we have
  \be
  \kappa(\tilde{\omega})=2\gamma_2 \tilde{\omega}
  \ee
with $\gamma_2=\frac{\sqrt{\pi}\Gamma(\frac74)}{3\Gamma(\frac54)}$.
As ${\omega}<z_0^{-1}$, both the retarded and advanced Green's functions become real-valued. This implies that there exists the threshold time scale $z_0^{-1}$, after which the particle dynamics become non-dissipative and depend on $z_0$. Thus, the strongly coupled field induces strikingly different effects on the particle compared to those given by the free field theory \cite{JT}.  In this late-time limit, the retarded and advanced Green's functions can be approximated by
\bea
 G^{(0)}_R(\omega)&\simeq & -\frac{R^2T_0}{z_0^3}\left(\tilde \omega + \frac{\tilde\omega^2}{\rho_m}- \rho_m^2 \tilde \omega^3 +\mathcal{O} (\tilde\omega^4)  \right)\, ,\\
 G^{(0)}_A(\omega) &\simeq & -\frac{R^2T_0}{z_0^3}\left(-\tilde \omega + \frac{\tilde\omega^2}{\rho_m}+ \rho_m^2 \tilde \omega^3 +\mathcal{O} (\tilde\omega^4) \right) \, .
\eea
Then the correlators become
\bea
  G^{(0)RR}(\omega)=G^{(0)LL}(\omega)&\simeq & \frac12(G^{(0)}_R(\omega)+G^{(0)}_A(\omega))+\frac{1+e^{4\gamma_2\tilde{\omega}}}{1-e^{4\gamma_2\tilde{\omega}}}\frac12(G^{(0)}_R(\omega)-G^{(0)}_A(\omega))\\
  &\simeq& -M_{z_0} \omega^2+\frac{R^2 T_0}{ 2 \gamma_2} \frac{1}{z_0^3} +\mathcal{O} (\omega^4)\,,
  \eea
and also
  \bea
  G^{(0)RL}(\omega)=G^{(0)LR}(\omega)&\simeq & \frac{e^{2\gamma_2\tilde{\omega}}}{1-e^{4\gamma_2\tilde{\omega}}} 
(G^{(0)}_A(\omega)-G^{(0)}_R(\omega))\\
  &\simeq& -\frac{R^2 T_0}{2 \gamma_2}\left( \frac{1}{z_0^3}-\frac{z_m^2}{z_0^3} \omega^2+\mathcal{O} (\omega^4)\right)\, .
  \eea
Thus, we find that the induced mass $M_{z_0}=M_0+ \Delta M_{z_0}$ now has a $z_0$ dependent minor correction $\Delta M_{z_0}=\frac{ R^2 T_0}{ 2 \gamma_2}\frac{z_m^2}{z_0^3}$. The cross correlators, $G^{(0)RL}(\omega)$ and $G^{(0)RL}(\omega)$, are dominated by the constant term of $1/z_0^3$, giving the decay rate in time as $t^{-1}$. Compared to the accelerating string case where the decay rate of the cross correlators is $\propto t^{-2}$, this potentially gives the higher disentanglement rate of the quark and anti-quark pair.
We will explore it in our future work.

\section{conclusion}\label{conclude}
In this paper, we study the holographic dual of the EPR pair of the quark and anti-quark in the super Yang-Mills theory. We calculate the two-point field correlators exactly, which are described by the linearized perturbations in the worldsheet of the string with two ends either undergoing uniform acceleration or being static. We find that the causality between two ends crucially determines the nature of the induced dispersion relation of the particles anchoring at the endpoints and also the feature of the mutual correlation between fields at the positions of two particles. For the causally disconnected two ends, the induced dispersion relation of each particle is dissipative, and the mutual correlation between fields dies out in the late times in the power law of time as $t^{-2}$. In contrast, for the casually connected two ends, the dispersion relation becomes non-dissipative in the late times, and the mutual correlation of fields decays in the power law of time as $t^{-1}$. There are some interesting problems to further understand the entanglement dynamics between two particles from the mutual field correlators. For example, we plan to explore how the time evolution of the entanglement entropy of the Brownian particles anchoring at two ends of a static string depends on the distance $z_0$ between two particles using the method in the paper \cite{Lee_19}.

Another interesting problem is considering the effects of quantum field fluctuations on the entanglement entropy between the entangled quark and anti-quark as found in \cite{Jensen_13} with the accelerating string solution. Compared to the weak coupling analysis in \cite{SY1,SY2}, we would obtain some insights into the time evolution of entanglement in strongly coupled theories. In particular, it is of interest to know whether or not the \textit{sudden death} of the entanglement entropy also occurs in strongly coupled theory. If so, since the entanglement is geometrically realized as a fundamental string connecting quark-antiquark pair, the time evolution may involve some nonperturbative effects leading to the change of the background solutions. For this purpose, we should extend the formula of the holographic influence functional in \cite{Lee_15} and \cite{Son_03} for finding the dynamics of a pair of the entangled particles.

In \cite{Maldacena_01}, the exponential decay of the correlation between two entangled boundary regions in the eternal AdS black hole implies the loss of information about the black hole behind the horizon. In the case of the accelerating string, the induced worldsheet background also has the causal structure the same as an eternal AdS black hole. However, the decay of the correlation between two boundary particles is expected as they couple to the super Yang-Mills fields, and the information can leak into the environments. If the ER=EPR conjecture also works in this case, the disentanglement of two particles can lead to the breakdown of the wormhole geometry, which is beyond the probe string approach we consider in this paper.  We do not have a solution for this problem yet. How this solution can be related to the black hole information problem is also an exciting problem to pursue.

\section*{Acknowledgments}

The work of S.~K. was supported in part by MOST109--2811--M--007--558
and MOST109--2112--M--007--018--MY2. The work of D.S. L. was supported in part by MOST-110-2112-M-259 -003. The work of C.P. Y. was supported in part by MOST-110-2112-M-259 -004 and MOST-109-2112-M-259 -006.

\end{document}